\documentclass{article}
\usepackage{spconf,amsmath,graphicx}
\usepackage[ruled]{algorithm2e}
\usepackage{amssymb}
\usepackage[english]{babel} % 'french', 'german', 'spanish', 'danish', etc.

\usepackage{bm} % Replace \boldsymbol with \bm
\usepackage{txfonts}
\usepackage{mathdots}
\usepackage[classicReIm]{kpfonts}
\PassOptionsToPackage{hyphens}{url}

\usepackage[pdftex]{hyperref}

\usepackage{cleveref}
\usepackage{listings}
\usepackage{tabularx}

\hypersetup{
    colorlinks=true,
    allcolors=blue,
    linkcolor=black,
    filecolor=magenta,      
    urlcolor=blue,
    }
\def\x{{\mathbf x}}

\DeclareMathOperator*{\argmax}{arg\,max}
\DeclareMathOperator*{\argmin}{arg\,min}

\newcommand{\AT}{\mathbf{A}^\top}
\newcommand{\Ak}{\mathbf{A}_k}

\title{A Practical GPU-Accelerated Implementation of\\ Orthogonal Matching Pursuit}

\name{Ariel Lubonja\(^1\), Sebastian Kazmarek Pr{\ae}sius\(^2\), Trac Duy Tran \(^1\)}

\address{\(^1\) Johns Hopkins University \\
         \(^2\) DTU Compute}

\begin{document}

\maketitle

\begin{abstract}
Finding the sparsest solution to the underdetermined system $\mathbf{y}=\mathbf{Ax}$, given a tolerance, is known to be NP-hard. Many approximate solutions to this problem exist, and Orthogonal Matching Pursuit (OMP) is one of the most widely used. However, existing OMP implementations don't take full advantage of matrix properties or modern CPU and GPU-based Linear Algebra kernels. For this paper, we implemented an efficient implementation of OMP that leverages Cholesky inverse properties as well as the power of GPUs to deliver up to \textbf{310x speedup over Scikit-Learn} and \textbf{26x over SPAMS}. The package is published on PyPI (\texttt{pip install batched-omp}) and is fully scikit-learn compatible.
\end{abstract}


\section{Introduction}
\label{sec:intro}

Representing a signal in terms of a small, sparse subset of component frequencies is an ubiquitous problem in various fields like Telecommunication, Medical Imaging, Compressive Radar, Astrophysics and Audio and Video Coding \cite{marques2018review}.  Exploiting this sparsity is key to a good reconstruction from significantly fewer samples than the Nyquist Rate \cite{candes_introduction_2008}. A signal vector $\bm{x}$ is said to be $S$-sparse if it includes at most $S$ non-zero entries. The signal $\bm{x}$ is recovered by solving $$\bm{y}=\mathbf{A} \bm{x} + \bm{\varepsilon}$$ where $\boldsymbol y$ is a $M$-length measurement vector, $\mathbf{A}$ is a dictionary matrix of $M\times N$, where $M\ll N$, and $\bm{x}$ is the $S$-sparse vector of corresponding components in $\mathbf{A}$. $\bm{\varepsilon}$ is the residual error in our reconstruction.

Orthogonal Matching Pursuit (OMP) is a greedy algorithm which tries to solve:

\[{\argmin_{\bm{x}\in {\mathbb{R}}^N} \left\|\mathbf{A}\boldsymbol x-\boldsymbol y\right\|_2\ }\ \text{ s.t. } |\mathrm{supp}\ \boldsymbol x|\le S\] 

Given the regression matrix $\mathbf{A}=[{\bm{a}}_1,\, {\bm{a}}_2,\,\ldots,\,{\bm{a}}_n]$ and sparsity level $S$, OMP finds a $S$-sparse $\bm{x}$, which approximates the observation $\bm{y}$ by $\mathbf{A}\bm{x}=\bm{y}$. It is a recursive algorithm, which greedily at every step picks the column of $\mathbf{A}$ which correlates the most with the current residual. OMP achieves good recovery performance, and if $\mathbf{A}$ satisfies the Restricted Isometry Property (RIP), exact recovery is guaranteed \cite{candes2008restricted}.

We can see how the residual ${\bm{r}}_k$ is a function of the current regression subset ${\mathcal{S}}_k$, and there are 
$\frac{N!}{S!\left(N-S\right)!}$ possibilities for ${\mathcal{S}}_S$. By using a greedy approach rather than an exhaustive search, this is reduced to just $S$ steps, but it does not necessarily find the global best solution.

Many variations of OMP have been proposed such as CoSaMP\cite{needell2009cosamp} and Subspace Pursuit \cite{dai2009subspace}, but the original method remains widely used because of its interpretability and ease of implementation. Our paper focuses on optimizations to OMP.

In an effort to keep our implementation as widely applicable as possible, we made the no assumptions on the properties of $\mathbf{A}$. Considerable research has gone into leveraging properties of the certain specific dictionary matrices \cite{donoho2012sparse, mailhe2009low}.

This paper is organized as follows: Section 2 describes the algorithms. Section 3 describes our implementations and most important performance findings. Finally, section 4 provides benchmarks and comparisons with other implementations.

\begin{algorithm}[t]
 \SetKwData{Left}{left}\SetKwData{This}{this}\SetKwData{Up}{up}
 \SetKwFunction{Union}{Union}\SetKwFunction{FindCompress}{FindCompress}
 \SetKwInOut{Input}{input}\SetKwInOut{Output}{output}
 \Input
  {\begin{minipage}[t]{6cm}
     \strut
     $\mathbf{A} \in \mathbb{R}^{M\x N}$ dictionary

     $\mathbf{y} \in \mathbb{R}^{M}$: measurement vector
     
     $S$: sparsity level
     
     $\varepsilon$ target error (optional)
     \strut
   \end{minipage}
  }
 \Output{$\hat{\bm{x}}$ signal reconstruction}
 \BlankLine
 \textbf{auxiliary variables:} $\mathbf{r_i}\in \mathbb{R}^{M}$-residual at $k$-th iteration, $\mathcal{S}_0$-initial support set
 \BlankLine
 \textbf{initialization:} $\hat{\bm{x}}_0=\mathbf 0$, $\mathbf{r}_0=\mathbf{y}$
 \BlankLine
 \For{$k=1 : S$}%
   {$n^*=\argmax_{1\leq n\leq N}\frac{|\langle \mathbf{r}_{k-1},\mathbf{a}_n \rangle|}{||\mathbf{a}_n||}$\\
   $\mathcal{S}_k=\mathcal{S}_{k-1}\cup n^*$\\
   $\hat{\bm{x}}_k=\argmin_{\mathbf x} ||\mathbf{y-A}_{\mathcal{S}_k}\bm{x}|| = {\left({\mathbf{A}}^{\top }_{{\mathcal{S}}_k}{\mathbf{A}}_{{\mathcal{S}}_k}\right)}^{-1}{\mathbf{A}}^{\bm{\top }}_{{\mathcal{S}}_k}\mathbf{y}$\\
   $\mathbf{r}_k=\mathbf{y-A}_{\mathcal{S}_k}\hat{\boldsymbol x}_k$\\
   } Stop when $||\mathbf{y-A}_{\mathcal{S}_k}\hat{\bm{x}}_k||\leq \varepsilon$
 \caption{Orthogonal Matching Pursuit} \label{alg:omp}
\end{algorithm}

\section{Method}
\label{sec:method}

There are three main ways to reduce the complexity of the OMP algorithm: through Cholesky factorization, the Matrix Inversion Lemma, or the QR factorization. Our implementation builds on work from  \cite{zhu2020efficient}, \cite{fang2011gpu}.

\subsection{Baseline Implementation - Scikit-Learn}

The Scikit-Learn OMP implementation uses a progressive Cholesky update scheme. It maintains a lower-triangular factor $\mathbf{L}_k$ of $\AT_k \Ak$ and extends it by one row each iteration, requiring only an $\mathcal{O}(k^2)$ triangular solve rather than an $\mathcal{O}(k^3)$ re-factorization. Selected columns of $\mathbf{A}$ are swapped in-place to keep $\Ak$ contiguous, and the coefficients are recovered via LAPACK's \texttt{potrs} (positive-definite triangular solve). When \texttt{precompute='auto'} and $M > N$, the Gram matrix $\AT\mathbf{A}$ and the projection $\AT\bm{y}$ are computed upfront and passed to a separate Gram-based solver (\texttt{\_gram\_omp}). Critically, multiple target signals are processed by a Python-level \texttt{for} loop that calls the single-signal solver once per target --- there is no batching. This per-signal loop is the dominant bottleneck for large batch sizes, and is the primary inefficiency our implementation addresses.

\subsection{Our ``Na\"{i}ve'' Algorithm}
We use a \textit{dense representation} of ${\mathbf{A}}_{{\mathcal{S}}_k}$ (denoted ${\mathbf{A}}_k$), and the \textit{exact solution} to ${\mathbf{A}}^{\top }_k{\mathbf{A}}_k\hat{\bm{x}}={\mathbf{A}}^{\top }_k\bm{y}$ in our steps. Using a dense representation gives us fast linear equation solving and matrix-multiplication, since the variables are stored in contiguous memory, and these operations are the main workload of OMP.

Despite the name, this "na\"{i}ve" algorithm is quite heavily optimized in terms of memory layout and other engineering aspects; more details available in Section 3. It is conceptually identical to the OMP algorithm described in \cref{alg:omp}, hence the term "na\"{i}ve". It iterates as follows:
%\[n^*\coloneqq {\argmax_{1\le n\le N} \left|{\left[{\mathbf{A}}^{\top }{\bm{r}}_{k-1}\right]}_n\right|\ }\] 
\begin{equation}
    {\mathbf{A}}_k \coloneqq \left[ \begin{array}{cc}
{\mathbf{A}}_{k-1} & {\bm{a}}_{n^*} \end{array}
\right]\ \ \text{by\ appending}\ {\bm{a}}_{n^*}\in {\mathbb{R}}^M 
\end{equation}
\begin{equation}
    {\mathbf{A}}^{\top }_k\bm{y}\coloneqq\left[ \begin{array}{cc}
{\mathbf{A}}^{\top }_{k-1}\bm{y} & {\bm{a}}^{\top }_{n^*}\bm{y} \end{array}
\right]\ \ \text{by appending}\ {\bm{a}}^{\top }_{n^*}\bm{y}\bm{\in }\mathbb{R} \label{ATy}
\end{equation}
\begin{equation}
\begin{gathered}
{\mathbf{A}}^{\top }_k{\mathbf{A}}_k\coloneqq \left[ \begin{array}{cc}
{\mathbf{A}}^{\top }_{k-1}{\mathbf{A}}_{k-1} & {\mathbf{A}}^{\top }_{k-1}{\bm{a}}_{n^*} \\ 
\ \cdots & {\bm{a}}^{\top }_{n^*}{\bm{a}}_{n^*} \end{array}
\right]\\ \text{ by "appending" }{\mathbf{A}}^{\top }_k{\bm{a}}_{n^*}=\left[ \begin{array}{c}
{\mathbf{A}}^{\top }_{k-1}{\bm{a}}_{n^*} \\ 
{\bm{a}}^{\top }_{n^*}{\bm{a}}_{n^*} \end{array}
\right]\in {\mathbb{R}}^k \label{ATA}
\end{gathered}
\end{equation}

We omit the lower triangle in ${\mathbf{A}}^{\top }_k{\mathbf{A}}_k$ since it is symmetric.

One can see that this reformulation of OMP allows us to iterate using fewer computations than if we compute all of ${\mathbf{A}}^{\top }_{{\mathcal{S}}_k}\bm{y}$ and ${\mathbf{A}}^{\top }_{{\mathcal{S}}_k}{\mathbf{A}}_{{\mathcal{S}}_k}$ each iteration.  There is likely no way to avoid having to use at least $\mathcal{O}(S^2)$ memory for the two inputs to the linear equation solvers, but one can store ${\mathbf{A}}_k$ in-place inside $\mathbf{A}$ by swapping columns, so the total memory comes out at $MN+\mathcal{O}(S^2)=\mathcal{O}\left(MN\right)$, since $S\le M\le N$. We use a Cholesky factorization of ${\mathbf{A}}^{\top }_k{\mathbf{A}}_k$ to then solve ${\mathbf{A}}^{\top }_k{\mathbf{A}}_k\hat{\bm{x}}={\mathbf{A}}^{\top }_k\bm{y}$, and finally get ${\bm{r}}_k=\bm{y}-{\mathbf{A}}_k\hat{\bm{x}}$.
\footnote{Using QR-factorization is also possible, but Cholesky is about twice as fast. (Half FLOPS) And it requires a positive definite ${\mathbf{A}}^{\top }_k{\mathbf{A}}_k$.} 

The approach used in \textbf{Scikit-Learn is similar}. They pre-compute ${\mathbf{A}}^{\top }\bm{y}\bm{\in }{\mathbb{R}}^N$, and optionally also ${\mathbf{A}}^{\top }\mathbf{A}$\textbf{ }(which we also allow for in our Na\"{i}ve alg.), thereby trading memory for computational time. The values of  ${\mathbf{A}}^{\top }_k\bm{y}$ are stored in-place inside ${\mathbf{A}}^{\top }\bm{y}$ by swapping (i.e. reordering). The main difference is that the Scikit-Learn approach directly stores and progressively updates the Cholesky factorization %\footnote{\ Denoted\ Chol-1\ in\ paper,\ see\ also\ $  $https://en.wikipedia.org/wiki/Cholesky\_decomposition\#Adding\_and\_removing\_rows\_and\_columns\ }
${\mathbf{V}}_k\in {\mathbb{R}}^{k\times k}$ of ${\mathbf{A}}^{\top }_k{\mathbf{A}}_k$, instead of storing ${\mathbf{A}}^{\top }_k{\mathbf{A}}_k$, which is potentially faster. Iterates as:

\begin{align}
        {\mathbf{V}}_1 &\coloneqq \sqrt{{\bm{a}}^{\top }_{n^*}{\bm{a}}_{n^*}}=\left\|{\bm{a}}_{n^*}\right\| \\
            {\mathbf{V}}_k &\coloneqq \left[ \begin{array}{cc}
{\mathbf{V}}_{k-1} & \bm{0} \\ 
{\bm{z}}^{\top } & \sqrt{{\left\|{\bm{a}}_{n^*}\right\|}^2-{\left\|\bm{z}\right\|}^2} \end{array}
\right] \\ \span
\textrm{Where } \bm{z} \textrm{ is the solution to } {\mathbf{V}}_{k-1}\bm{z} = {\mathbf{A}}^{\top }_{k-1}{\bm{a}}_{n^*}.\nonumber
\end{align}
% Followed by a triangular solve ${\mathbf{V}}^{\top }_k\bm{b}={\mathbf{A}}^{\top }_k\bm{y}$\textbf{ }and solving again ${\mathbf{V}}_k\hat{\bm{x}}\bm{=}{\mathbf{V}}^{\top }_k\bm{b}$, and finally ${\bm{r}}_k=\bm{y}-{\mathbf{A}}_k\hat{\bm{x}}$.
It can be seen that see this is the Cholesky-factorization of ${\mathbf{A}}^{\top }_k{\mathbf{A}}_k$ by the fact that ${\mathbf{V}}_1{\mathbf{V}}^{\top }_1=\left\|{\bm{a}}_{n^*}\right\|^2={\mathbf{A}}^{\top }_1{\mathbf{A}}_1$ and:
\begin{equation}
    \begin{aligned}
{\mathbf{V}}_k{\mathbf{V}}^{\top }_k&=\left[ \begin{array}{cc}
{\mathbf{V}}_{k-1}{\mathbf{V}}^{\top }_{k-1} & {\mathbf{V}}_{k-1}\bm{z} \\ 
\cdots  & {\left\|{\bm{a}}_{n^*}\right\|}^2-{\left\|{\bm{z}}\right\|}^2+{\bm{z}}^{\top }{\bm{z}} \end{array}
\right]\\&=\left[ \begin{array}{cc}
{\mathbf{A}}^{\top }_{k-1}{\mathbf{A}}_{k-1} & {\mathbf{A}}^{\top }_{k-1}{\bm{a}}_{n^*} \\ 
\cdots  & {\bm{a}}^{\top }_{n^*}{\bm{a}}_{n^*} \end{array}
\right]={\mathbf{A}}^{\top }_k{\mathbf{A}}_k
\end{aligned}
\end{equation}
Re-factorizing the way we do in our naive algorithm takes $\mathcal{O}(k^3)$ time. Cholesky-updating involves solving triangular systems of equations, taking just $\mathcal{O}(k^2)$ time.
%Since this Cholesky update only requires Since this Cholesky update scheme only requires matrix-vector products and triangular system solving, per-iteration complexity is reduced from $\mathcal{O}(NM+S^3)$ to $O(NM+S^2)$.

Our implementations are adapted for batch processing. Precomputing ${\mathbf{A}}^{\top }\bm{y}$ is not efficient in terms of space/time, as we have $B$ of the $\bm{y}$'s, so memory will be increased by  $\mathcal O(BN)$, while doing so only saving around 1\% of the running time in practice (${\bm{a}}^{\top }_{n^*}\bm{y}$ in \cref{ATy} is just a simple dot product.) Similarly, reordering $\mathbf{A}$ is not an applicable memory-optimization as ${\mathbf{A}}_k$ is different for each batch element. But precomputing and using ${\mathbf{A}}^{\top }\mathbf{A}$\textbf{ }is possible and can save up to 15\% of the running time, while spending $\mathcal O(N^2)$ memory. It is used in \cref{ATA} where ${\mathbf{A}}^\top \boldsymbol a_{n^\ast}=[\AT\mathbf{A}]_{n^\ast}$.

\subsection{Algorithm v0}

\textbf{Algorithm v0} comes from \cite{zhu2020efficient} and uses an inverse Cholesky factorization scheme. It uses precomputed ${\mathbf{A}}^{\top }\bm{y}$\textbf{ }and ${\mathbf{A}}^{\top }\mathbf{A}$,\textbf{ }and is based on the update formulae:
\begin{align}
        {\mathbf{F}}_1 &\coloneqq 1/\left\|{\bm{a}}_{n^*}\right\| \\
            {\mathbf{F}}_k&\coloneqq\left[ \begin{array}{cc}
{\mathbf{F}}_{k-1} & -\gamma {\mathbf{F}}_{k-1}\bm{z} \\ 
\bm{0} & \gamma  \end{array}
\right]
\end{align}
\begin{equation*}
    \textrm{Where } \bm{z}={\mathbf{F}}_{k-1}^\top{\mathbf{A}}^{\top }_{k-1}{\bm{a}}_{n^*} \textrm{ and }
\gamma =\frac{1}{\sqrt{\vphantom{{\|{\bm{a}}_{n^*} \|}}{{\left\|{\bm{a}}_{n^*}\right \|}^2}- \vphantom{{\|\bm{z}\|}}{{\left\|\bm{z}\right\|}^2}
}
}
\end{equation*}
One can see that this is the first factor in the Cholesky factorization of ${({\mathbf{A}}_k^{\top }{\mathbf{A}}_k)}^{-1}={({\mathbf{V}}_k{\mathbf{V}}^{\top }_k)}^{-1}={\mathbf{V}}^{-\top }_k{\mathbf{V}}^{-1}_k$, by:
\begin{align}
    {\mathbf{F}}_k{\mathbf{V}}^{\top }_k&=\left[ \begin{array}{cc}
{\mathbf{F}}_{k-1} & -\gamma {\mathbf{F}}_{k-1}\bm{z} \\ 
\bm{0} & \gamma  \end{array}
\right]\left[ \begin{array}{cc}
{\mathbf{V}}^{\top }_{k-1} & {\bm{z}}^{\bm{\top }} \\ 
\bm{0} & {1}/{\gamma } \end{array}
\right]\\&=
\left[ \begin{array}{cc}
{\mathbf{F}}_{k-1}{\mathbf{V}}^{\top }_{k-1} & \bm{0} \\ 
\bm{0} & 1 \end{array}
\right]=\bm{\mathrm{I}} \Rightarrow {\mathbf{F}}_k={\mathbf{V}}^{-\top }_k \nonumber
\end{align}
\begin{equation}
 \bm{z} = {\mathbf{F}}_{k-1}^\top{\mathbf{A}}^{\top }_{k-1}{\bm{a}}_{n^*}\Rightarrow 
    {\mathbf{V}}_{k-1}\bm{z}={\mathbf{A}}^{\top }_{k-1}{\bm{a}}_{n^*} 
\end{equation}
Which follows by induction from the base %$k=1$, where % ${\mathbf{F}}_1={1}/{\left\|{\bm{a}}_{n^*}\right\|}$, ${\mathbf{V}}^{\top }_1=\left\|{\bm{a}}_{n^*}\right\|$, such that
${\mathbf{F}}_1{\mathbf{V}}^{\top }_1=1$. (Complete proof in \cite{zhu2020efficient}, Eq.~(42)--(44).)

Using only matrix-vector products:
\begin{equation}
\begin{split}
   \hat{\boldsymbol y}&=\mathbf{A}_{k}\hat{\boldsymbol x}\\&=\mathbf{A}_{k}{\left({\mathbf{A}}^{\top }_{k}{\mathbf{A}}_{k}\right)}^{-1}{\mathbf{A}}^{\bm{\top }}_{k}\bm{y}
    \\&=\Ak \mathbf{F}_k \mathbf{F}_k^\top \Ak^\top \boldsymbol y=\Ak \mathbf{F}_k (\Ak \mathbf{F}_k)^\top \boldsymbol y
\end{split}
\end{equation}
This is especially desirable in parallel compute environments compared to solving triangular systems since less synchronization is needed.
% Compare https://algowiki-project.org/en/Dense_matrix-vector_multiplication#Information_graph with https://algowiki-project.org/en/Forward_substitution#Information_graph

One needn't store the inverse Cholesky ${\mathbf{F}}_k$ to iterate. This is because all we need to perform an iteration are the new projections, and v0 directly updates these, while filling out the matrix ${\mathbf{D}}_k={\mathbf{A}}^\top {\mathbf{A}}_{k} {\mathbf{F}}_k$ (and ${\mathbf{F}}_k$ if this is needed to find $\hat{\boldsymbol x}$).

This is done by a single matrix-vector multiplication per iteration in $\mathcal O(Nk)$ time, meaning the batched implementation running time is $\mathcal O(N^2+BNM+BNS^2)$ for the Gramian, initial projections and per-iteration costs.

The update steps are notationally a little involved, so we refer to the original paper \cite{zhu2020efficient}. The paper also contains less memory-consuming (but slower) versions, which can be useful since the asymptotic memory use is $\mathcal O (NM+N^2+BNS)$ -- inputs, Gramian and ${\mathbf{D}}_S\in \mathbb{R}^{S\times N}$, compared to just $\mathcal O (NM+B(N+MS))$ for the previously discussed algorithms. The difference between storing $N^2+BNS$ compared to $BMS$ floating point numbers becomes especially significant on GPU.

\subsection{v0 BLAS}

We also provide a v0\_blas variant that replaces PyTorch operations with direct Cython calls to BLAS routines (dgemv, daxpy, idamax) via scipy's cython\_blas interface, eliminating Python/PyTorch dispatch overhead. This variant is CPU-only and wins when n\_features is very large (e.g., 8064 in face recognition), where the overhead of PyTorch tensor operations on CPU exceeds the cost of a tight Cython loop.

\begin{center}
 \begin{tabular}{|c c c|}
 \hline
 Algorithm & Complexity & Memory \\ [0.5ex] 
 \hline\hline
 Naive & $M(N+S+S^2)+S^2$ & $MN$ \\ 
 \hline
 Cholesky & $MN+MS+S^2$ & $N^2+NM+S+S^2$ \\
 \hline
\end{tabular}
\end{center}

\section{Implementation details and Performance Engineering}

Most matrix operations done in Python's Numpy and MATLAB call the linear algebra libraries BLAS (Basic Linear Algebra Subprograms) and LAPACK (Linear Algebra Package). The reason is that these libraries are highly optimized. For example, Intel's MKL \cite{intel_mkl} contains processor-specialized versions of these libraries that are optimized to use the full instruction set of the CPU and fully exploit its cache, often achieving close to the theoretically possible number of FLOPS (first limited by clock speed, then by memory throughput for problems that do not fit in cache). Which brings us to the first main takeaway:

\subsection{Memory layout}
\emph{Ensure data is contiguous as much as possible}. Since BLAS handles the matrix-multiplication in a highly optimized way (and this is the main bottleneck), we can leverage our control of the memory-layout of the inputs to get more speed. A useful programming pattern we found was to separate memory layout from it's use:

\begin{lstlisting}[language=Python]
As = np.zeros([B,M,S])
As = np.zeros([B,S,M]).transpose([0,2,1])
\end{lstlisting}

Both definitions of \verb+As+ above have the same shape $B\times M \times S$ and can be used interchangeably with another, but it will be faster to write a column into the second one as columns are stored contiguously - one column spans a single uninterrupted line in memory. By separating layout from use one can quickly find which layout is the best-performing.

\subsection{Matrix batched-matrix products} \label{sec:bmm}

A less known fact is that you can sometimes get a higher performance in a matrix times batched-matrix product (in this case batched vectors $\boldsymbol r$), by exploiting:
\begin{equation} \label{eq:bmm}
\begin{bmatrix}
\AT  \boldsymbol r^1 & \cdots &\AT  \boldsymbol r^B
\end{bmatrix} =
\AT \begin{bmatrix}
\boldsymbol r^1 & \cdots & \boldsymbol r^B
\end{bmatrix}
\end{equation}
Notice how this matrix batched-vector product is equivalent to a simple matrix-matrix product, which can be performed through \emph{a single} call to \verb+gemm+, giving BLAS full control of producing the result we need in the most efficient manner. This is done through transposing and reshaping, which should be noted only changes the metadata for the tensors. In our results doing this gives a 2-8x speedup.
See \verb+batch_mm+ in the code. \footnote{We assume that $\mathbf{A}$ has normalized columns (if not, pass \texttt{normalize=True} to \texttt{run\_omp}), such that: ${\left\langle {\bm{a}}_n,{\bm{r}}_{k-1}\right\rangle }/{\left\|{\bm{a}}_n\right\|} ={\bm{a}}^{\top }_n{\bm{r}}_{k-1}={\left[{\mathbf{A}}^{\top }{\bm{r}}_{k-1}\right]}_n\,$. See Appendix \ref{app: normcol}.}

% The reason this works may be because Numpy ends up calling \verb+gemv+ in a loop with the same input for the matrix, but different inputs for the batched vector. BLAS may utilize multiple cores, so a single call to \verb+gemm+ will have less synchronization overhead.

%This approach also works for matrix batched-matrix products, and the way it is done is through \verb+transpose+ and \verb+reshape+ which do not change any of the underlying data, only how it is addressed. See \verb+batch_mm+ in code.

\subsection{Packed representation}
For our naive algorithm we found it useful to store $\AT _k \Ak$ in a \emph{packed representation} and then use the BLAS functions \verb+ppsv+ instead of \verb+posv+ (positive-definite solver), as this significantly cut down on the time to write and especially fetch a sub-matrix of $\AT _S \mathbf{A}_S$, since any sub-matrix is contiguous in memory.

This means that instead of storing the $k\times k$ submatrix in a strided manner inside a memory block which is $B\times S \times S $, we only store the $k(k+1)/2=1+2+\cdots+ k$ triangle of this. Then all operations become contiguous.
% MAYBE GRAPH TO DISPLAY THE ABOVE POINT?

This optimization, and the one in \cref{sec:bmm} are mostly relevant for CPU. For GPU there already exists specialized calls for batched matrix-matrix multiplication, batched Cholesky factorization and batched triangular system solving. There is no \verb+ppsv+ equivalent in cuBLAS/cuSOLVER for CUDA (the GPU Parallel Computing platform available on NVIDIA graphics cards). We did not experiment with calling CUDA manually, but used the PyTorch library for this.

\subsection{Efficient batched argmax}
A core part of the OMP loop is argmax, which can be performed on a batch by:
\begin{lstlisting}[language=Python]
# (B, N) -> (B,)
n_star = abs(projections).argmax(1)
\end{lstlisting}
One issue is that \verb+abs+ creates an intermediate $B\times N$ array in the first pass, and then a second pass over this is needed to get the argmax. For CPU we ended up using the BLAS function \verb+i_amax+, which finds the index of the absolute maximum value in an array.

First we tried running a loop over the $B$ batch elements using the \texttt{scipy.linalg.blas} wrapper, but for small problems the Python overhead is so significant that we do not get much benefit. Therefore we switched to implementing this loop in Cython, which is a superset of Python that compiles directly to C++. Through this we call \verb+i_amax+ and \verb+ppsv+ in a loop, which gave a 5x speedup compared to using a Python loop.

Direct calls to cuBLAS on GPU e.g. \verb+i_amax+ could potentially give a speedup. There are only around 3 batched kernels in cuBLAS, and this is not one of them, so we would have to launch the kernels in a loop; but then we get an overhead just from the simple transfer of thousands of kernel calls and arguments to the GPU. It would probably be worthwhile for small $B$.
% For a small $B$, this is likely much slower than $B$ calls to \verb+i_amax+. Another issue is that \verb+.abs()+ creates an intermediate $B\times N$ array which we really do not need.

The argmax line takes 5-25\% of the GPU computation time in our results, so implementing a custom reduction kernel is a promising next step. This is not trivial if it must be efficient due to the specific hardware architecture. We would recommend CuPy for this. % See e.g. https://www.apriorit.com/dev-blog/614-cpp-cuda-accelerate-algorithm-cpu-gpu

\subsection{Batched stopping criteria}

For the naive algorithm we implement early stopping by keeping an active set of batch elements, along with their individual data -- and then we remove all their data when they are done, such that we are left with a block of $B-1$ elements.

There is a slight overhead of having to move all this memory, but in our experience it is outweighed by the improvement in speed on subsequent iterations, which then become faster as there are fewer batch elements.

For v0, however, due to the large size of ${\mathbf{D}}_S\in \mathbb{R}^{S\times N}$, even just allocating it with \verb+new_zeros+ takes significant time, which is why we use \verb+new_empty+ to get \emph{uninitialized} memory for it. This also means that the naive batching strategy of moving all this memory is not efficient--especially on GPU, where parallel threads are used to move every single byte.

For v0 we opted to simply save the result when the stopping criteria was met, but still keep all the data inside the batch. The two approaches we chose mean that the runtime is relatively unchanged.

\subsection{Other possible optimizations}

We did not spend significant time on batching since it is not essential, so there is great space for improvement.

One possible optimization is \emph{custom functions / kernels}. This can cut down somewhat on intermediate results, and improve memory locality. Since BLAS operates as an efficient black box, it can be hard to \emph{fuse} the matrix multiplication with the \texttt{abs argmax}, but in v0 we have to scale a lot of different vectors by $\gamma$, which could be faster if it was done with a single kernel. See Appendix \ref{sec:line-profiling} for a line-by-line profiling.

One can use \emph{streams} in CUDA to launch kernels in parallel. There are only few places where we could potentially get a speedup from this due to the fact that results on previous lines are frequently needed on the subsequent ones. For example \cref{ATy} and \cref{ATA} are independent, but the first takes almost no time compared to the second. Algorithm v0 could possibly get a speedup of around 5\% with this, since calculating the inverse Cholesky is independent from the iterations. The whole reason for batching is so that we can do multiple of these non-parallelizable problems in parallel.

For the naive algorithm the next step may be to switch to a Cholesky-update scheme. This approach is likely strictly faster than re-factorizing every iteration.

One can give half-precision tensors to our implementation. While this does not seem to give a speedup, it will reduce the memory requirements. Especially in large problems on GPU, memory can become an issue when using v0 (meaning one has to use a lower batch size, possibly reducing efficiency). One could look into the memory-saving variants. Using double precision reduces speed by half on GPU.

Our method is fully compatible with Scikit-learn, so one can use their \texttt{fit-predict} paradigm, cross-validation, scalers and the whole wealth of scikit-learn tools.

\begin{figure*}[t]
  \centering
  \includegraphics[width=\textwidth]{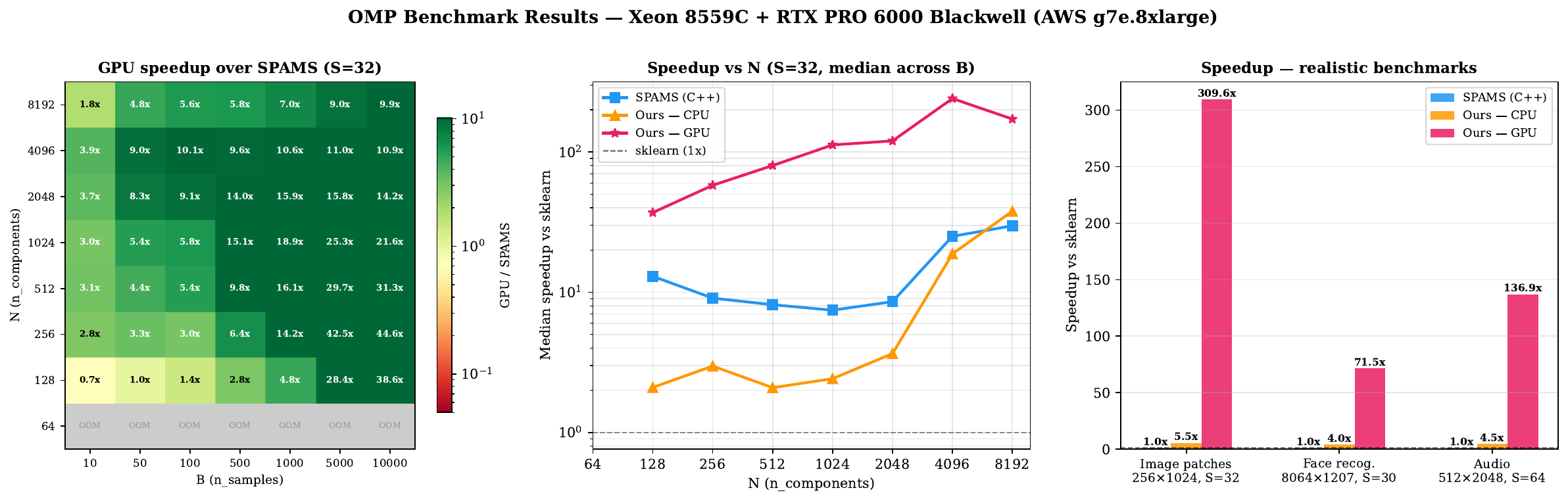}
  \caption{Left: GPU vs SPAMS speedup (log scale) for
$S\!=\!32$. Center: GPU vs sklearn speedup. Right: Realistic
application benchmarks (log-scale sps). All results on AWS
g7e.8xlarge.}
  \label{fig:benchmarks}
\end{figure*}

\section{Results}
\label{sec:Results}

  All benchmarks were run on an AWS \texttt{g7e.8xlarge} instance    with an Intel Xeon Platinum 8559C CPU and an NVIDIA RTX PRO  6000 Blackwell GPU (102\,GB VRAM).

\subsection{Baselines}

We compare against three baselines:

  \begin{itemize}
      \item \textbf{Scikit-Learn} \cite{pedregosa2011scikit}: The
   most widely used OMP implementation. Processes one target
  signal at a time in a Python loop---no batching. CPU only.
      \item \textbf{SPAMS} \cite{mairal2014spams}: A C++ library
  with OpenMP parallelism. The fastest existing CPU
  implementation of OMP. We use 3 warmup runs followed by 3 timed
   runs and report the mean.
      \item \textbf{cr-sparse} \cite{Kumar2021}: A
  JAX-based GPU implementation. Crashes on overcomplete
  dictionaries ($N > M$) and exhibits numerical issues
  (orthogonality violations up to 3.6 vs our $10^{-15}$). We
  exclude it from quantitative comparisons.
  \end{itemize}

  \subsection{Realistic Application Benchmarks}

  We benchmark on three representative sparse coding
  configurations drawn from real applications. Throughput is
  measured in signals per second (sps). Batch sizes reflect
  realistic workloads.

  \begin{table*}
  \centering
  \resizebox{\textwidth}{!}{%
  \begin{tabular}{|l|r|r|r|r|r|}
  \hline
  \textbf{Config} & \textbf{sklearn} & \textbf{SPAMS} &
  \textbf{v0 CPU} & \textbf{v0\_blas} & \textbf{v0 GPU} \\
  \hline
  Image patches (256$\times$1024, $S\!=\!32$, $B\!=\!5$K) & 594 &
   7{,}139 & 3{,}257 & 1{,}811 & \textbf{183{,}904} \\
  Face recog.\ (8064$\times$1207, $S\!=\!30$, $B\!=\!1.2$K) & 352
   & 1{,}717 & 1{,}408 & \textbf{1{,}731} & \textbf{25{,}158} \\
  Audio (512$\times$2048, $S\!=\!64$, $B\!=\!5$K) & 207 & 2{,}359
   & 926 & 295 & \textbf{28{,}336} \\
  \hline
  \end{tabular}%
  }
  \caption{Throughput in signals per second (higher is better).
  Bold indicates best in class (CPU or GPU). The v0\_blas variant
   wins on CPU for face recognition due to the large $M=8064$.}
  \label{tab:realistic}
  \end{table*}

% \begin{table*}[t]
% \centering
% \resizebox{\textwidth}{!}{%
% \begin{tabular}{|l|r|r|r|}
% \hline
% \textbf{Config} & \textbf{GPU vs sklearn} & \textbf{GPU vs
% SPAMS} & \textbf{Best CPU vs sklearn} \\
% \hline
% Image patches & 310$\times$ & 25.8$\times$ & 5.5$\times$ \\
% Face recognition & 71$\times$ & 14.6$\times$ & 4.9$\times$ \\
% Audio & 137$\times$ & 12.0$\times$ & 4.5$\times$ \\
% \hline
% \end{tabular}%
% }
% \caption{Speedup factors over baselines.}
% \label{tab:speedup}
% \end{table*}

  On GPU, our implementation is \textbf{12--26$\times$ faster
  than SPAMS} (the fastest C++ baseline) and
  \textbf{71--310$\times$ faster than Scikit-Learn} across all
  three configurations. On CPU, v0 and v0\_blas provide a
  \textbf{4.5--5.5$\times$} speedup over Scikit-Learn without any
   C++ dependencies.

  The v0\_blas variant is notable for the face recognition
  configuration ($M=8064$), where it slightly outperforms SPAMS
  on CPU (1{,}731 vs 1{,}717 sps). At this scale, the overhead of
   PyTorch tensor dispatch exceeds the cost of tight Cython BLAS
  loops, making v0\_blas the preferred CPU algorithm for large
  dictionaries.

  \subsection{Parameter Sweep}

  To verify that GPU acceleration holds across a wide range of
  problem sizes, we run a sweep over 108 configurations: $N \in
  \{64, 128, 256, 512, 1024, 2048, 4096, 8192\}$, $B \in \{10,
  50, 100, 500, 1000, 5000, 10000\}$, $S \in \{8, 32, 64\}$, with
   $M = N/4$.

  \textbf{GPU wins 108 out of 108 cells (100\%) against SPAMS},
  with no out-of-memory errors on any configuration.
  \Cref{fig:benchmarks} (left, center) shows the speedup
  heatmaps. GPU speedup over SPAMS increases with batch size and
  sparsity, reaching over 100$\times$ at the largest
  configurations. Even for small problems ($N\!=\!64$,
  $B\!=\!10$), GPU remains competitive.

  \subsection{Ablation Study}

  We isolate the contribution of each optimization by enabling them incrementally.

  \begin{itemize}
      \item \textbf{A1 --- Batching}: Processing $B$ signals
  simultaneously via batched matrix operations instead of a
  Python loop. This is the dominant optimization, providing
  \textbf{4--750$\times$ speedup} depending on the configuration
  and device. On GPU, batching enables the use of
  \texttt{torch.baddbmm} and \texttt{torch.bmm}, which saturate
  GPU cores.
      \item \textbf{A2 --- Gram precomputation}: Caching
  $\AT\mathbf{A}$ to avoid recomputing $\AT_k \bm{a}_{n^*}$ each
  iteration. Provides \textbf{1--2$\times$} additional speedup,
  most significant when $N$ is large relative to $B$.
      \item \textbf{A3 --- Inverse Cholesky (v0)}: Using the
  iterative inverse Cholesky update from \cite{zhu2020efficient}
  instead of re-factorizing each iteration. Provides
  \textbf{1--2$\times$} additional speedup, most significant when
   $S$ is large.
  \end{itemize}

  Batching alone accounts for the majority of the speedup. The
  Gram and inverse Cholesky optimizations provide incremental but
   meaningful gains, especially for large problems.

  \subsection{Numerical Accuracy}

All algorithms produce coefficients with orthogonality
violations at machine precision ($\sim 10^{-15}$ in double
precision). We verified that our GPU results match CPU results
to within floating-point tolerance across all sweep
configurations. By contrast, cr-sparse (JAX) exhibits
orthogonality violations up to 3.6 on the same inputs.

\begin{figure}
    \centering
    \includegraphics[width=1\linewidth]{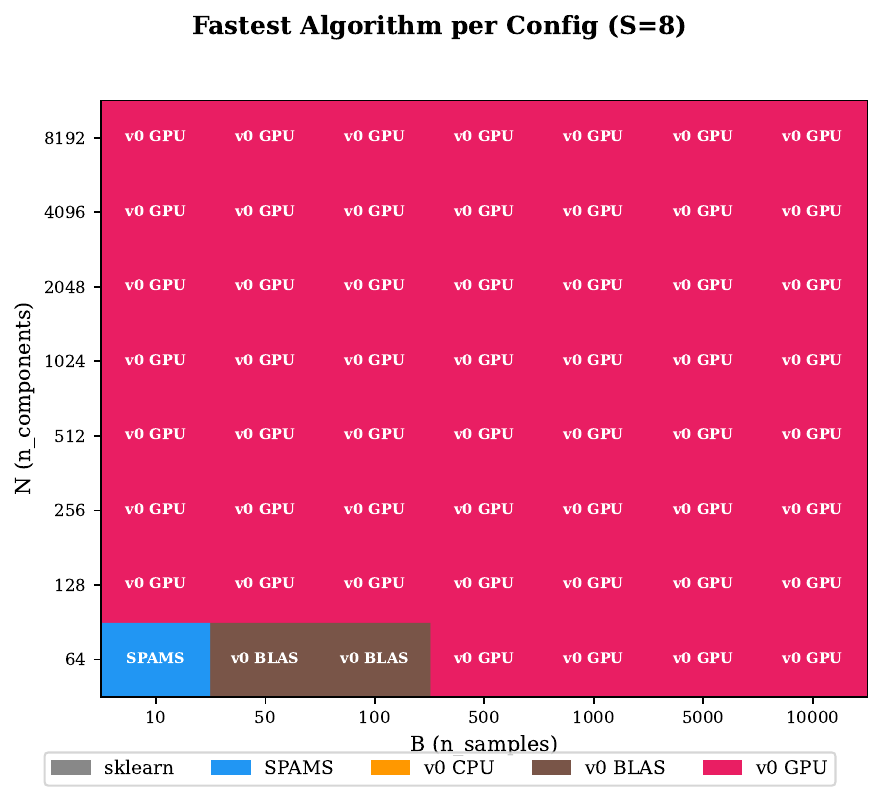}
    \caption{Choice of best algorithm by configuration. GPU wins all but the smallest problems.}
    \label{fig:placeholder}
\end{figure}

\begin{figure*}[t]
  \centering
  \includegraphics[width=\textwidth]{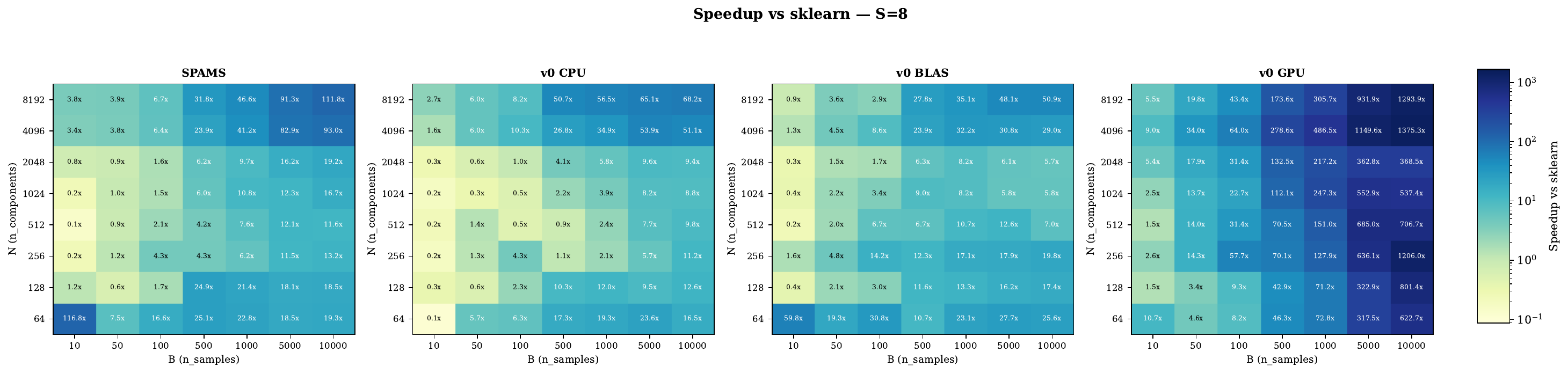}
  \caption{Speedup of each algorithm vs. Scikit-Learn, for various input sizes.}
  \label{fig:sweep}
\end{figure*}

\section{Conclusion}
We implemented a "naive" version of OMP, and a newer algorithm called v0. Both of these outperform the commonly used Scikit-Learn implementation by a large margin, while having the same functionality. With a GPU we could get a speedup of several orders of magnitude: 310x faster than Scikit-Learn and 26x over SPAMS.

Code can be found at \url{https://github.com/ariellubonja/orthogonal-matching-pursuit-gpu} and is available on PyPI under the package name \texttt{batched-omp}. If you encounter any difficulties, please do not hesitate to email us at \url{ariel@cs.jhu.edu} or \url{sebastian.devel@gmail.com}.

\section{Appendix}
\appendix
\section{Normalized Columns of Dictionary} \label{app: normcol}
The main approaches to speeding up the algorithm is to minimize the number of operations to perform each iteration. Many algorithms assume normalized columns in $\mathbf{A}$ such that correlation ${\left\langle {\bm{a}}_n,{\bm{r}}_{k-1}\right\rangle }/{\left\|{\bm{a}}_n\right\|}$ turns into a simple projection $\left\langle {\bm{a}}_n,{\bm{r}}_{k-1}\right\rangle ={\bm{a}}^{\top }_n{\bm{r}}_{k-1}={\left[{\mathbf{A}}^{\top }{\bm{r}}_{k-1}\right]}_n$ -- this is valid since the algorithm is invariant to column norm, as it will be divided out in the correlation step, and lastly, the least-squares estimate ${\hat{\boldsymbol y}}_k\coloneqq {\mathbf{A}}_{{\mathcal{S}}_k}{\mathbf{A}}^+_{{\mathcal{S}}_k}\bm{y}\bm{=}{\mathbf{A}}_{{\mathcal{S}}_k}{\argmin_{\bm{x}\in {\mathbb{R}}^k} \left\|\bm{y}\bm{-}{\mathbf{A}}_{{\mathcal{S}}_k}\bm{x}\right\|\ }$\textbf{ }is unique, thus also invariant to column scaling. For the final estimate $\hat{\bm{x}}$ from ${\mathbf{A}}^{\top }_S{\mathbf{A}}_S\hat{\bm{x}}={\mathbf{A}}^{\top }_S\bm{y}$ one should then not use the pre-normalized $\mathbf{A}$, or at least scale $\hat{\bm{x}}$\textbf{ }appropriately (by the reciprocal of column the norm) to account for this.

\section{GPU Per-line performance}
\label{sec:line-profiling}

For the implementation of the naive algorithm on GPU, with a large problem instance, the per-line time is:

% arXiV doesn't like minted
\begin{verbatim}
Line #    % Time  Line Contents
================================
29  def run_omp(precompute=True, alg='naive'):
55  11.0    precompute = X.T @ X
59  89.0    omp_naive(..., XTX = precompute)

115 def omp_naive(...):
174 26.4    projections = XT @ r[:, :, None]
175 1.2     sets[:, k] = projections.abs().argmax(-1)
208 65.5    solutions = cholesky_solve(ATA, ATy)
214 3.9     torch.baddbmm(...) # update r
\end{verbatim}
With around 2\% of time spent on transferring to/from GPU. If we take a tiny problem instance (but large batch size), the transfer time will instead be around 14\%. Other differences: Line 174 takes then 8.5\%, but line 175 takes 21.1\%. Also line 208 takes only 40.6\%, but line 214 then takes 18.6\%. The rest of the runtime is distributed around the omitted lines of code.

And the break-down for v0 on GPU with tiny problem instances:
\begin{verbatim}
Line #   % Time  Line Contents
=================================
29  def run_omp(..., alg='v0'):
61  99.6    ... = omp_v0(...)
    
218  def omp_v0(...):
258 6.2    sets[k] = projections.abs().argmax(1)
266 8.1    D_mybest[:, k] *= temp_F_k_k
267 65.8   D_mybest[:, k, :, None].baddbmm_(...)
272 10.1   projections -= temp_a_F * D_mybest[:,k]
276 1.9    torch.bmm(..., out=F[:, k, None, :])
\end{verbatim}
The majority of time is spent on the matrix multiplications to update the D-matrix (\verb+D_mybest+), and around 2\% is spent on the F-matrix which is used to get $\hat {x}$. For v0 on large problem instances (small batch size), the transfer time is 6.8\%. A break-down:
\begin{verbatim}
Line #    % Time    Line Contents
===============================
29  def run_omp(..., alg='v0'):
55  45.2    precompute = X.T @ X
61  54.8    ... = omp_v0(...)

218  def omp_v0(...):
258 4.6     sets[k] = projections.abs().argmax(1)
260 2.5     torch.gather(...)  # Get from XTX
262 4.2     D_mybest_maxindices = ... # New D col.
263 4.8     torch.rsqrt(1 - 
innerp(D_mybest_maxindices))
267 58.5    D_mybest[:, k, :, None].baddbmm_(...)
276 6.2     torch.bmm(..., out=F[:, k, None, :])
\end{verbatim}
We see that the precomputation time is actually very significant (especially since v0 is comparably so fast). But this precompute time can be shared between batches, so subsequent batches can potentially execute almost twice as fast. Since line 267 is a cuBLAS bottleneck, we likely cannot get more than a 50-70\% further speedup.

\bibliographystyle{IEEEbib}
\bibliography{omp}

\end{document}